\documentclass[12pt,preprint]{aastex}
\def \apj {ApJ}
\def \apjl {ApJ Lett}

\def \aap {A\&A}
\def \araa {ARA\&A}

\def\arcmin{\hbox{$^\prime$}}
\def\arcsec{\hbox{$^{\prime\prime}$}}

\begin{document}
%
%
\shorttitle{GRB 050319 : Wind to ISM transition}
\shortauthors{Kamble et al.}
\title{Observations of the Optical Afterglow of GRB 050319 : Wind to ISM transition in view}

\author{
Atish Kamble\altaffilmark{1},
L. Resmi\altaffilmark{1,2},
Kuntal Misra\altaffilmark{3}
}
\altaffiltext{1} {Raman Research Institute, Bangalore - 560 080, India}
\altaffiltext{2} {Joint Astronomy Programme, Indian Institute of Science, Bangalore - 560 012, India}
\altaffiltext{3} {Aryabhatta Research Institute of Observational Sciences, Manora Peak, Nainital - 263 129, India}
~~~~~~~~~\emph{\copyright 2007. The American Astronomical Society. All rights reserved.}
\begin{abstract}
The collapse of a massive star is believed to be the most probable progenitor 
of a long GRB. Such a star is expected to modify its environment by stellar wind.
The effect of such a circum-stellar wind medium is expected to be seen 
in the evolution of a GRB afterglow, but has so far not been conclusively found.
We claim that a signature of wind to constant density medium transition
of circum-burst medium is visible in the afterglow of GRB 050319.
Along with the optical observations of the afterglow of GRB 050319
we present a model for the multiband afterglow of GRB 050319. We show that
the break seen in optical light curve at $\sim$ 0.02 day could be explained 
as being due to wind to constant density medium transition of circum-burst medium,
in which case, to our knowledge, this could be the first ever detection of such 
a transition at any given frequency band. Detection of such a transition
could also serve as a confirmation of massive star collapse scenario
for GRB progenitors, independent of supernova signatures.
\end{abstract}


\section{Introduction}

One of the long standing questions in astrophysics is the progenitors
of Gamma Ray Bursts (GRBs). Collapse of a massive star is one of the most favoured
progenitors of the long GRBs. Evidence for a massive star being 
a GRB progenitor may be obtained in two different ways, both using 
observations of GRB afterglows. 
{\it (1) Supernova (SN) component underlying the GRB afterglow} :
A few of the nearby GRB afterglows have shown the temporal and spectroscopic 
signature of an underlying SN. see e.g., \citep{030329_00} 
{\it (2) Evolution of GRB afterglow in the stellar wind medium} : 
Massive stars modify the density profile of the circum-stellar medium
due to the powerful winds they drive during their life time. For a constant mass loss rate 
and constant wind velocity the circum-burst medium assumes a density profile $\rho \propto r^{-2}$
as compared to $\rho =$ constant in the absence of stellar wind.
Evolution of the GRB afterglow light curves is significantly different in these two cases of
density profiles \citep{Wijers1999, Chevalier2000}. Attempts
to look for the signatures of such a wind-modified circum-burst density profile in the light curves 
of GRB afterglows have not been conclusive so far. In the case of GRB 050904 \citet{050904_Gendre}
finds that the early x-ray afterglow suggests a wind like density profile of the circum-burst
medium while the late optical afterglow was consistent with evolution in a constant density medium.
Hence, they conjecture that a transition between these two types of density profiles 
would have taken place somewhere in between. However, this transition was not directly observed
in the light curve of any given band. We show that the afterglow of GRB 050319 could be explained 
as being due to the transition of the circum-burst density profile from wind-like
to constant density.\\

GRB 050319 was detected by the Burst Alert Telescope (BAT) instrument of the SWIFT satellite
on 2005 March 19, 09:31:18.44 UT \citep{GCN3117,GCN3119}.
However, \citet{Cusumano}, using the re-analysis of the BAT data, pointed out that 
{\em Swift} was slewing during the GRB onset and the BAT trigger was switched off. 
The GRB was recognised about 135 seconds after its actual onset.
The total duration of the GRB ($\rm T_{90}$) was thus 149.7 s \citep{Cusumano}
instead of $10 \pm 2$ s \citep{GCN3117,GCN3119}. The burst fluence in 15-350 keV band 
within the $\rm T_{90}$ duration is estimated to be 1.6 $\times$ 10$^{-6}$ erg cm$^{-2}$.
The photon index of the time-averaged single power law spectrum is 2.1 $\pm$ 0.2.
{\em Swift} XRT and UVOT located a bright source at $\alpha$ = $10\rm^{h} 16\rm^{m} 48\rm^{s}$ 
and $\delta$ = $+43^{\circ} 32^{\arcmin} 47^{\arcsec}$ (J2000) which was later confirmed 
by \citet{GCN3116} with ROTSE-IIIb. \citet{GCN3136} obtained the spectra 
of the afterglow of GRB 050319 on 2005 March 20 and the redshift of the afterglow 
was measured to be z = 3.24. At this redshift, the gamma ray isotropic 
equivalent energy released during the burst is $3.7 \times 10^{52}$ erg for a flat universe with
$\rm \Omega_{m} = 0.3, \Omega_{\Lambda} = 0.7 ~and~ H_{0} = 70 ~km~s^{-1} ~Mpc^{-1}$.\\

\section{Observations and Data Reduction} 
Optical CCD observations of the afterglow of GRB 050319 were carried 
out in Johnson BV and Cousins RI filters using the 104-cm Sampurnanand Telescope 
of ARIES, Nainital with regular specifications of the CCD camera and using standard 
observation procedures of bias subtraction and flat fielding. For details see \citet{060124_Misra}.

The BVRI magnitudes of the optical transient (OT) obtained were calibrated 
differentially using secondary stars numbered 8, 9, 10, 11 and 13 in the list of \citet{GCN3454}.
The magnitudes derived in this way are given in Table ~\ref{tab:data}.
Photometric magnitudes available in the literature by \citet{050319_Raptor,Quimby,050319_Mason,Huang}
were converted to the present photometric scales using the five secondary stars mentioned above.
\begin{table}[h]
\medskip
\begin{center}
\begin{tabular} {cccc}\hline
\hline
Date (UT)&Magnitude&Exposure Time&Passband\\
2005 March&(mag)&(s)&\\
\hline
&&&\\
 19.7512&21.02$\pm$0.14&2$\times$900&B\\
&&&\\
19.7457&20.23$\pm$0.20&600&V\\
19.8558&20.60$\pm$0.21&600&V\\
&&&\\
19.6949&19.46$\pm$0.16&300&R\\
19.6999&19.25$\pm$0.10&300&R\\
19.7215&19.98$\pm$0.11&300&R\\
19.7539&20.06$\pm$0.13&300&R\\
19.7874&20.03$\pm$0.16&300&R\\
&&&\\
19.7129&19.59$\pm$0.19&300&I\\
19.7587&19.66$\pm$0.25&300&I\\
&&&\\
\hline
\end{tabular}
\end{center}
\caption{CCD BVRI broad band optical observations of the GRB 050319 afterglow using the 104-cm 
Sampurnanand Telescope at ARIES, Nainital.}\label{tab:data}
\end{table} 
\section{Light curves of GRB 050319}

Along with our own observations we have used observations reported in
the literature to study the light curves of GRB 050319 afterglow.
The x-ray afterglow was observed by {\em Swift} XRT starting from 
$\sim 220$s to 28 days \citep{Cusumano} after the burst.
Attempts to observe the afterglow 
at radio wavebands resulted in upper limits \citep{GCN3127,GCN3132,GCN3153}. 
The optical afterglow was observed by \citet{050319_Raptor,Quimby,050319_Mason}
resulting in a coverage from a few seconds to $\sim 4$ days
after the burst.

To construct the optical light curve we have corrected the observed magnitudes
for the standard Galactic extinction law given by \citet{Mathis}.  The galactic
extinction in the direction of GRB 050319 is estimated to be E(B-V) = 0.011
mag from the smoothed reddening map provided by \citet{Schlegel}.
The effective wavelength and normalization given by \citet{Bessel} were used to
convert the magnitudes to fluxes in $\mu$Jy.\\

Most of the GRB afterglow light curves are well characterized
by a broken power law of the form
$\rm F = F_{0}\{(t/t_{b})^{\alpha_{1}s} + (t/t_{b})^{\alpha_{2}s}\}^{-1/s}$
where $\alpha_{1}$ and $\alpha_{2}$ are the afterglow flux decay indices
before and after the break time ($\rm t_{b}$), respectively.
$\rm F_{0}$ is the flux normalization and `s' is a smoothening parameter which
controls the sharpness of the break. Most known GRB afterglows have 
$\alpha_{2} > \alpha_{1}$ i.e. the decay becomes steeper
after the break. Interestingly, the optical afterglow light curve of GRB 050319
shows steeper to flatter decay with a break at $\sim 0.02$ day. 
This behavior of light curve decay is difficult to explain within the standard 
afterglow models.
The x-ray and optical light curves also show some variability superimposed on 
the power law decay. The x-ray light curve shows a break near $\sim$ 0.3 day.
We quantify the various characteristics of the afterglow 
light curves as summarized below.\\

\begin{itemize}

\item[1.] The x-ray afterglow of GRB 050319 shows a very rapid decay before 
$\sim 0.005$ day ($\sim 384$ s) after the GRB. The decay then flattens before 
steepening again at $\sim 0.3$ day. 
\citet{Cusumano} has characterised the afterglow into three separate temporal
evolutions in x-ray bands : 
$\rm \alpha_{X1} = 5.53 \pm 0.67 ~(\Delta t < 384~s)$;
$\rm \alpha_{X2} = 0.54 \pm 0.04 ~(384 ~s < \Delta t < 0.3 day)$;
$\rm \alpha_{X3} = 1.14 \pm 0.2  ~(\Delta t > 0.3 day)$;
\item[2.] The B, V and R band light curve also show a rapid decline during 
the early phase ($\rm \Delta t < 0.02$ day) which flattens at later epochs. 
Thus, the afterglow can be separated into two separate temporal 
evolutions in R band 
$\rm \alpha_{R1} = 1.09 \pm 0.03 ~(\Delta t < 0.02 day)$;
$\rm \alpha_{R2} = 0.51 \pm 0.03 ~(0.02 day < \Delta t < 10.0 day)$;\\
And for B, V and I bands we measure following $\alpha$ s :
$\rm \alpha_{B1} = 1.46 \pm 0.26 ~(\Delta t < 0.02 day)$;
$\rm \alpha_{B2} = 0.33 \pm 0.05 ~(0.02 day < \Delta t < 1.0 day)$;
$\rm \alpha_{V1} = 0.90 \pm 0.05 ~(\Delta t < 0.02 day)$;
$\rm \alpha_{V2} = 0.50 \pm 0.04 ~(0.02 day < \Delta t < 1.0 day)$;
$\rm \alpha_{I2} = 0.59 \pm 0.13 ~(0.02 day < \Delta t < 1.0 day)$;
The average decay index of the optical light curve, at early and late epochs, 
is then $1.15 \pm 0.27 $ and $ 0.48 \pm 0.15$ respectively.

\end{itemize}

\section{GRB 050319 afterglow : wind or homogeneous density profile ?}

The breaks seen in x-ray light curve (at $\sim 384$ s and at $\sim 0.3$ day) 
are not accompanied by simultaneous breaks in optical wavebands.
Similarly, the break seen in optical band has no simultaneous counterpart 
in the x-ray light curve. Also, the sense of slope change, 
i.e. $\alpha_{2} < \alpha_{1}$, as seen in optical waveband is contrary 
to the predictions of the fireball model \citep{Sari01, Sari02} which 
expects $\alpha_{1} < \alpha_{2}$.

We propose a different model to explain the afterglow 
of GRB 050319 as being due to a transition of the circum-burst medium 
density profile that the explosion generated shock wave is interacting with. 
We propose that the observed change from steep to flat decay of the 
optical afterglow of GRB 050319 at 0.02 day is due to the change in 
the density profile of the circum-burst medium from wind modified ($\rho \propto r^{-2}$) 
to the constant density medium ($\rho =$ constant). 
The break in the light curve occurs when the shock front interacts with 
the boundary between the two density profiles. Below we describe and 
reproduce various features of the GRB 050319 afterglow using this
model of `wind to constant density medium transition'.

The early steep decay of x-ray afterglows ($\alpha \sim$ 3 to 5) 
as is seen in the case of GRB 050319 are now seen routinely 
in most of the GRBs \citep{Steep00} and has become 
a canonical feature of the GRB x-ray afterglows.
In the case of GRB 050319 \citet{Cusumano} conjecture that the early steep decay 
emission could be low energy tail of the GRB prompt emission. We exclude this 
early emission from the rest of our discussion and we will restrict ourselves 
to the rest of the x-ray light curve.

The radiation spectrum of GRB afterglows exhibits a power law spectrum
characterised by three break frequencies - the self absorption
frequency $\rm \nu_{a}$, the peak frequency $\nu_{\rm m}$
corresponding to the lower cutoff in the electron energy distribution
($n(\gamma) \propto \gamma^{-p}$, $\gamma > \gamma_{m}$), and
the synchrotron cooling frequency $\nu_{\rm c}$.  The flux $F_{\rm m}$
at $\nu_{\rm m}$ provides the normalisation of the spectrum
\citep{Sari01}.

The photon index ($\Gamma$) of the afterglow and the electron energy
distribution index $p$ are related in any given spectral regime
($\Gamma - 1 = p/2$ if $\rm \nu_{c} < \nu$ and
$\Gamma - 1 = (p - 1)/2$ if $\rm \nu < \nu_{c}$).
The corresponding temporal decay index $\alpha$ would be $(3p-2)/4$ 
and $3(p-1)/4$ respectively before the jet break and would equal $p$ in both spectral
regimes after the jet break, according to the standard fireball model
for an afterglow expanding in a homogeneous interstellar medium (ISM).
For the shock wave expanding into the wind density profile 
the corresponding $\alpha$ would be $(3p-2)/4$ and $(3p-1)/4$
respectively before the jet break and $p$ after the jet break.

In the present case, the observed values of the photon index
($\Gamma = 1.69 \pm 0.06$) and temporal decay index 
($\alpha_{\rm X} = 0.54 \pm 0.04$) of the x-ray afterglow
are consistent with the spectral regime $\rm \nu_{X} > \nu_{c}$ 
and $p=1.5$. The observed decay indices of optical light curve
are also consistent with the inferred value of $p$, the spectral regime
$\rm \nu_{\rm opt} < \nu_{c}$ and wind-constant density transition at 0.02 day.
As discussed above, the expected temporal decay index of the x-ray 
afterglow, $\alpha_{\rm X} = (3p-2)/4$, is the same for wind and homogeneous ISM 
density profiles. The absence of a break in the x-ray afterglow 
light curve simultaneous with the optical break makes the multiband afterglow
features consistent with the proposed transition of the circum-burst medium 
density profile from wind to constant density.
In Figure 1. we compare the predictions of our model of wind to constant 
density transition of the circum-burst density profile with 
multiband observations of GRB 050319 afterglow.
A detailed list of the best fit spectral parameters can be found in Table ~\ref{tab:fitpara}.
The observed B band light curve is systematically lower than that predicted by
the model which could be due to the $Lyman ~\alpha$ absorption at $z = 3.24$ 
appearing in the observer's B band as suggested by \citet{Huang}.
The steepening of the x-ray light curve at $\Delta t \sim 0.3$ day could be due 
to jet break \citep{Cusumano} but unfortunately the variability in 
R band light curve and insufficient sampling of the data in B and V bands
after $\sim 1.0$ day makes it difficult to verify the achromaticity of the break.

Given the above model spectral parameters we find that the broadband behavior 
of the afterglow is very well explained. However, we are restricting ourselves 
to the overall behavior of the afterglow and hence do not attempt in our model 
to reproduce the variations seen in the optical light curve. 
The reason for these variations could be density inhomogenities in 
the circum-burst medium.\\

\subsection{Physical Parameters}
Four spectral parameters ($\rm \nu_{a} ,
\nu_{m} , \nu_{c}\ and\  F_{peak}$) are related to four physical parameters
viz n (no density of the constant density circum-burst medium) or 
$A_{\ast}$ (defined as $\rho(r) = 5 \times 10^{11} A_{\ast} r^{-2}$ 
for wind density medium),
E (total energy content of the fireball), energy fraction in relativistic
electrons $\rm \epsilon_{e}$ and that in magnetic field $\epsilon_{B}$.
The typical value of self absorption frequency $\rm \nu_{a}$ lies in
radio-mm waves and hence is best estimated only if the afterglow is well
observed in these bands. Unfortunately, the afterglow of GRB 050319
was never detected at the radio band \citep{GCN3127,GCN3132,GCN3153}. 
Therefore we expressed the remaining three spectral parameters, 
known separately from parts of the light curves corresponding to the wind 
and the constant density circum-burst medium, in terms of A$_{\ast}$ and $n$,
respectively. Equating the kinetic energies estimated from two density profiles,
i.e. $E^{K}_{wind} = E^{K}_{ISM}$, we obtain a relation between A$_{\ast}$ and $n$ :
A$_{\ast} = 5.097 \times 10^{-3} ~n^{2/5}$. For a typical range of values of 
$n$ (1 to 100), estimated A$_{\ast}$ ranges from $5.097 \times 10^{-3}$
to $0.032$ which results in the range of $E^{K}_{iso}$ from $1.3 \times 10^{54}$ 
to $5.3 \times 10^{53}$ erg. All the estimated physical parameters 
are listed in Table ~\ref{tab:physpara}.

\section{Discussion}
\subsection{Signature of Wind Reverse Shock ?}
Morphology and evolution of the wind bubbles has been studied 
by \citet{Castor,Weaver}. A reverse shock forms at the surface where 
the stellar wind meets the surrounding ISM and it then propagates into the wind.
The free wind (upstream of the reverse shock) has a density profile $\rho \propto r^{-2}$
and the shocked wind (downstream of the reverse shock) has a constant density profile.
The effects of such a density transition on the afterglow of a GRB
has been studied by \citet{Peer}.
It could be this transition of the density profile that we are observing at 0.02 day
in the present case of GRB 050319. From the observations of long GRB afterglows,
it has been inferred that most of the GRBs occur in constant density environment
and the absence of wind signatures in the GRB afterglow
was surprising. Various ways which can bring the wind reverse shock closer 
to the exploding star have recently been proposed to resolve this mystery 
surrounding the absence of winds \citep{vanMarle,Eldridge}.
In the case of GRB 050319, for a range of assumed values
of $n = 1 ~to~ 100$ we estimate the radius of the reverse shock to be
$R_{SW} \sim 0.5$ pc to $0.1$ pc comparable to the values obtained by \citet{Eldridge_2006}.
The constraint $\epsilon_{B} < 1$ puts a lower bound on density :
$n > 0.03$

\subsection{Implications for the models of GRB progenitors}

Our interpretation of the afterglow of GRB 050319 as being due to the 
wind-constant density transition supports the collapsar model of GRBs.
Detection of similar transitions in other afterglows have so far eluded us
perhaps because of the smaller size of the wind bubbles and the resultant 
early transition times. The present detection was made possible chiefly 
because of the quick follow up abilities of the robotic telescopes 
RAPTOR \citep{050319_Raptor} and ROTSE-III \citep{Quimby} coupled with 
those of {\em Swift} XRT \citep{Cusumano} and UVOT \citep{050319_Mason}. 
Time dilation due to cosmological redshift delays 
the occurrence of the transition in the observer's frame of reference and 
makes it favourable to detect such a transition in distant GRBs.
The robotic telescopes are now routinely detecting GRB afterglows
as early as a few minutes after the burst and with careful analysis 
of multiband observations of distant GRBs it should be possible to detect 
more examples of similar transition. For example, a probable detection of 
similar density transition, though not in the same waveband, has been reported 
by \citet{050904_Gendre} in the case of GRB 050904 (z $\sim$ 6.3) 
where the transition time is assumed to be $\sim 1700$ s ($\sim$ 0.02 days) 
after the burst, similar to that for GRB 050319 afterglow in the present case.
\section{Summary}
We have modeled the multiband afterglow of GRB 050319, using our own optical observations 
and other observations available in the literature, as being due to 
the interaction of the relativistic blast wave 
with circum-burst medium which shows a transition of density profile from 
wind to constant density. 
Our conclusions can be summarised as follows :
\begin{enumerate}
\item We present BVRI band observations of GRB 050319 afterglow.
\item We showed that the unusual break in the light curves of optical afterglow 
at 0.02 day can be explained as being due to the transition of circum-burst density
profile from wind to constant density. The observed x-ray afterglow light curve 
without a simultaneous break is consistent with this interpretation.
The overall afterglow can be explained by using a relatively low value 
of electron energy distribution index $p$ which is also consistent with 
the x-ray spectral photon index.
\item The transition of the density profile could be due to the wind reverse
shock propagating into the stellar wind driven by the progenitor of GRB 050319.
We estimate radius of the wind reverse shock
to be $R_{SW} \sim 0.5 ~pc ~to~ 0.1 ~pc$ for assumed values of 
$n \sim 1 ~to~ 100$ cm$^{-3}$ respectively.
\end{enumerate}
\begin{figure}
\begin{center}
\hbox{
\includegraphics[height=15cm,angle=-90]{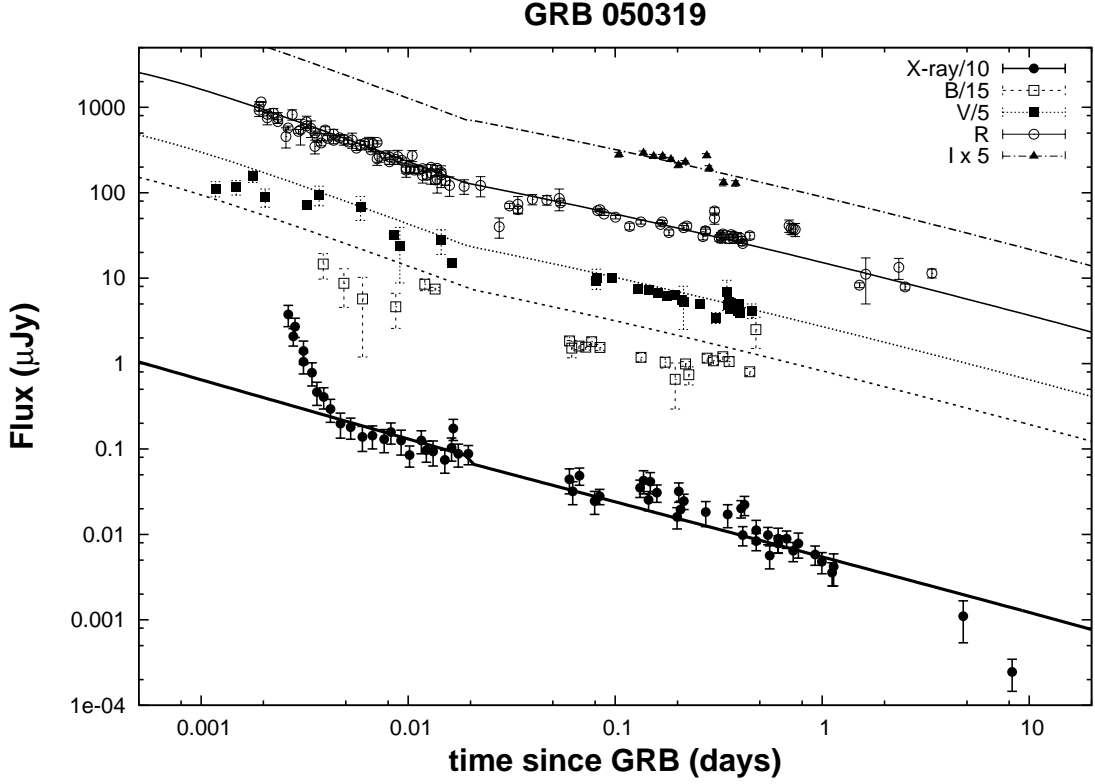}}
\caption{The afterglow light curves of GRB 050319. 
The solid lines represent a model in which the expanding fireball 
encounters the transition in density profile from the wind to constant density medium 
at 0.02 day. The best fit spectral parameters of this model are listed 
in Table ~\ref{tab:fitpara}}
\end{center}
\end{figure}
\begin{table}
\begin{center}
\begin{tabular}{|c|c|c|}
\hline
\hline
                        &      wind density medium       	&     constant density medium         	\\
\hline
$ \nu_{m} (Hz)$      & $1.6^{+0.9}_{-0.7}\times 10^{13}$    & $1.0^{+0.35}_{-0.5} \times 10^{12}$  \\
$ \nu_{c} (Hz)$      & $1.1^{+3.7}_{-0.4}\times 10^{15}$    & $2.1^{+0.6}_{-0.8} \times 10^{15}$         \\
$ F_{peak}(\mu Jy)$  & $2370 \pm 355$                   	& $1810^{+260}_{-170}$      		\\
$p$                 & $1.59 \pm 0.06$                       & $1.52 \pm 0.02$                 \\
\hline
$\chi^{2}_{dof} (dof)$&    \multicolumn{2}{c|}{1.4 (161)}       \\
\hline
\end{tabular}
\end{center}
\caption{The best fit spectral parameters for wind ($< 0.02$ day) and constant density ($> 0.02$ day) profile. 
All the parameters are fitted at 0.003 day after the burst.}\label{tab:fitpara}
\end{table}
\begin{table}
\begin{center}
\begin{tabular}{|c|c|c|c|c|}
\hline
\hline
Parameter		&    \multicolumn{2}{c|}{n = 1; A$_{\ast} = 5.097\times10^{-3}$}&\multicolumn{2}{c|}{n = 100; A$_{\ast} = 0.032$}\\
\hline
                        &      wind      	&     ISM         	&	wind		&	ISM	\\
\hline
$ E^{iso}_{54} $	& 	$1.3$      	&	$1.3$ 		& 	$0.53$		&	$0.53$    \\
$\epsilon_{e} $	& $4.4\times 10^{-3}$   &	$2.2\times 10^{-3}$& 	$1.1 \times 10^{-2}$	&	$5.4 \times 10^{-3}$    \\
$\epsilon_{B} $	& $0.14$    		& 	$1.2\times 10^{-3}$&	$0.01$    	& 	$ \sim 10^{-4}$           \\
\hline
\end{tabular}
\end{center}
\caption{Physical parameters estimated using the best fit spectral parameters mentioned in Table ~\ref{tab:fitpara}}\label{tab:physpara}
\end{table}
\section*{Acknowledgments}
We are thankful to D. Bhattacharya for critical comments and detailed discussions 
throughout this work and to Ram Sagar for his support during the observations.
We are thankful to the anonymous referee for constructive comments which 
has improved the $Letter$ significantly.


\clearpage
\newpage

\label{lastpage}
\end{document}